\documentclass[twocolumn,superscriptaddress,aps,preprintnumbers,amsmath,amssymb,prl,nofootinbib]{revtex4}

\usepackage{dcolumn}
\usepackage{graphicx}
\usepackage{epstopdf}
\usepackage{bm}
\usepackage{hyperref}
\usepackage{color}

\begin{document}

\title{
Hunting Axion Dark Matter with Protoplanetary Disk Polarimetry
}

\author{Tomohiro Fujita}
\affiliation{Department of Physics, Kyoto University, Kyoto, 606-8502, Japan}
\affiliation{D\'epartment de Physique Th\'eorique and Center for Astroparticle Physics, Universit\'e de Gen\`eve, Quai E.Ansermet 24, CH-1211 Gen\`eve 4, Switzerland}
\author{Ryo Tazaki}
\affiliation{Astronomical Institute, Tohoku University, 6-3 Aramaki, Aoba-ku, Sendai 980-8578, Japan}
\author{Kenji Toma}
\affiliation{Frontier Research Institute for Interdisciplinary Sciences, Tohoku University, Sendai 980-8578, Japan}
\affiliation{Astronomical Institute, Tohoku University, 6-3 Aramaki, Aoba-ku, Sendai 980-8578, Japan}

\begin{abstract}
We find that the polarimetric observations of protoplanetary disks are useful
to search for ultra-light axion dark matter. Axion dark matter predicts the
rotation of the linear polarization plane of propagating light, and protoplanetary
disks are ideal targets to observe it. We show that a recent observation puts the tightest constraint on the axion-photon coupling constant for axion mass $m\lesssim10^{-21}$eV.
\end{abstract}
\maketitle

%
%
%
\section{I. Introduction}

The identity of dark matter has been one of the most fundamental questions in physics and astronomy for decades. Among many dark matter candidates, axion or axion like particles (ALPs) are particularly studied for their strong motivations in particle physics.
Axion was originally proposed to solve the strong $CP$ problem by Peccei and Quinn~\cite{Peccei:1977hh}, while similar pseudo-scalar fields are ubiquitously predicted by string theory and they are called ALPs~\cite{Witten:1984dg} (see recent reviews~\cite{Dias:2014osa,Graham:2015ouw}). In this letter, we call both of them axion for short in the context of dark matter.
It is known that the axion field $\phi(t)$ should oscillate around its mass potential $\phi\sim \cos(mt)$ to behave as dark matter, where $m$ is the axion mass. 
Therefore, to verify the axion dark matter (ADM), the mass is an important parameter. Since its allowed region spans 28 orders of magnitude, $10^{-25}{\rm eV}\lesssim m\lesssim 10^3$eV~\cite{Hlozek:2014lca,Arias:2012az}, and the optimized experiment to discover the ADM depends on $m$, we need to select our target mass range to devise  a detection scheme.

Cosmological simulations of the cold dark matter dynamics tend to predict more cuspy halo profiles
and a larger number of low-mass halos than the observations on small scales~\cite{Weinberg:2013aya}.
These issues are called the core-cusp problem and the satellite problem, respectively, and they are under intense debate.
The possibility of dark matter with an extremely small mass $m\sim 10^{-22}$eV attracts attention as a solution for these issues~\cite{Hui:2016ltb,Hu:2000ke}.
Having such a tiny mass, the dark matter significantly shows its quantum nature even on astrophysical scales, and the quantum pressure from the uncertainty principle suppresses the smaller structure formation so that the predictions and the observations are consistent if $m\sim 10^{-22}$eV~\cite{Hu:2000ke,Schive:2014dra}.
The fit to the rotation curves of galaxies also favors the same mass~\cite{Bernal:2017oih}.
Therefore, the dark matters with such a small mass (called as fuzzy dark matter) have an enhanced motivation,
and the ADM is its ideal candidate, because it is natural for axion to have a small mass by virtue of the shift symmetry.
Note that a tension between the fuzzy dark matter scenario and observations of the Lyman-$\alpha$ forest was pointed out~\cite{Irsic:2017yje} which is still under discussion~\cite{Zhang:2017chj}.

Several previous works attempted to test the ultra-light ADM with $m\sim 10^{-22}$eV by using the coupling between axion and photon, $\mathcal{L}_{a\gamma}=-g_{a\gamma}\phi F_{\mu\nu}\tilde{F}^{\mu\nu}/4$.
This interaction predicts two different phenomena: One is the conversion between an axion particle and a photon under magnetic fields, and the other is the photon birefringence under axion background.
Exploiting the former phenomenon, axion helioscope~\cite{Anastassopoulos:2017ftl,Armengaud:2014gea}
and ``light shining through a wall'' experiments~\cite{Ehret:2010mh}
are searching for axion. Astrophysical observations of SN1987A~\cite{Payez:2014xsa}
and quasars~\cite{Payez:2012vf} were also studied to detect the signatures of the conversion.
Unfortunately, no signal is yet to be observed (see however \cite{Angus:2013sua,Meyer:2013pny,Kohri:2017ljt}). Since the conversion probability is controlled by the coupling constant $g_{a\gamma}$, the previous works put the upper bound, $g_{a\gamma}\lesssim 10^{-11}{\rm GeV}^{-1}$ for $m\lesssim 10^{-14}$eV. 
It should be noted that these astrophysical constraints suffer from the uncertainty
of cosmic magnetic fields, because the conversion probability also depends on their strength and structure.

In this Letter, we demonstrate a novel approach to search for ultra-light ADM with $m\gtrsim 10^{-22}$eV.
First, we consider not the axion-photon conversion but
the photon birefringence.
With the oscillating ADM background, the photon dispersion relation is modified
and the linear polarization plane of photon rotates, as we will see in Sec.~II.
Although this phenomenon has been known for a long time~\cite{Carroll:1989vb,Harari:1992ea},
the experiments for the ADM with birefringence were discussed only recently~\cite{DeRocco:2018jwe}. Second, we propose a new observed object suitable for the ultra-light ADM search, protoplanetary disk (PPD). 
PPD is a flattened gaseous object surrounding a young star, where planets are thought to be formed. Recent study of planet formation mechanism has been rapidly developed by intense observational efforts with a number of new instruments at optical, infrared, and radio wavebands. They revealed that PPDs are bright simply by scattering the central star's light at optical and near-infrared wavebands~\cite{Testi14}. This simple physics enables us to figure out the linear polarization pattern at the source, so that current high-quality polarimetric observations of PPDs can be used to diagnose the photon birefringence during its propagation.

With our new approach the best constraint on $g_{a\gamma}$ for $m\sim 10^{-22}$eV without the uncertainty of magnetic fields
is obtained, and we open a fresh new possibility of
PPD observations as dark matter search.

\section{II. Rotation of polarization plane}

In this section, we show that 
the linear polarization plane of a propagating photon rotates under 
the background of the ADM.
It can be shown that the equation of motion for the photon field (i.e. the vector potential) $\bm{A}(t,\bm x)$ gains additional terms from the coupling to the axion $\phi(t,\bm x)$ as
\begin{equation}
\ddot{\bm A} - \nabla^2 \bm{A} = 
g_{a\gamma}\left(\dot{\phi}\,\bm{\nabla}\times \bm{A}+\dot{\bm A}\times \bm{\nabla}\phi\right) \,,
\label{EoM A}
\end{equation}
where $g_{a\gamma}$ is the coupling constant of the photon-axion coupling $\mathcal{L}_{a\gamma}$ and dot denotes partial derivative with respect to time.
Here, we chose the temporal gauge $A_0 = 0$ and the Coulomb gauge $\bm{\nabla}\cdot \bm{A} = 0$. Note that the cosmic expansion is ignored because the time scale
in interest is much shorter than the Hubble time.

Considering that the correction from the photon-axion coupling is tiny, if any, we study the deviation from the normal plane wave solution of photon by neglecting the time evolution of its amplitude. 
We decomposes $\bm{A}$ into the circular polarization modes in the Fourier space,
\begin{equation}
\bm{A}(t, \bm{x}) = \sum_{p=\pm}\int\frac{{\rm d}^3 k}{(2\pi)^3}A^p_{k}\bm{e}^p(\hat{\bm{k}})e^{i\bm{k}\cdot\bm{x}
-i\int{\rm d}t\, \omega_p} \,,
\end{equation}
where  we introduced the time integral of $\omega_\pm$ instead of the plane wave solution $e^{-i\omega_{\pm} t }$, since $\omega_\pm$ depends on time for fixed $k$ as we will see below.
The circular polarization vector satisfies $\bm{e}^p(\hat{\bm{k}}) = \bm{e}^{p*}(-\hat{\bm{k}})$, $\bm{e}^{p}(\hat{\bm{k}})\cdot \bm{e}^{*p'}(\hat{\bm{k}}) = \delta^{pp'},$ and $i\bm{k}\times\bm{e}^\pm(\hat{\bm{k}}) = \pm k \bm{e}^\pm(\hat{\bm{k}})$.
Then, at the leading order of $g_{a\gamma}$, the dispersion relation is given by~\cite{Harari:1992ea}
\begin{align}
\omega_\pm &\simeq k\pm \delta \omega,
\\
\delta\omega &= -\frac{g_{a\gamma}}{2} \left[\dot{\phi}+\hat{\bm{k}}\cdot{\bm \nabla}\phi\right]=-\frac{g_{a\gamma}}{2}\frac{{\rm d} \phi}{{\rm d} t},
\end{align}
where we assume $k\gg |\delta \omega|$ and $k\gg\partial_t\ln|{\rm d}\phi/{\rm d}t|$, and ${\rm d}/{\rm d}t$ is the total derivative along the light path with $\partial\bm{x}/\partial t=\hat{\bm{k}}$.
Therefore under the ADM background, the phase velocities of the left/right-handed modes are different. 
This is because the parity symmetry is spontaneously broken by the axion field.
This photon birefringence caused by the axion background leads to the rotation of the linear polarization plane.

Provided that a photon propagating along the z-axis is linearly polarized into the x-direction at the initial time $t$, its polarization components can be decomposed into the circular ones,
\begin{align}
\left(\begin{array}{c}
1 \\
0 \\
\end{array}\right)
=\frac{1}{2}
\left(\begin{array}{c}
1 \\
i \\
\end{array}\right)
+
\frac{1}{2}
\left(\begin{array}{c}
1 \\
-i \\
\end{array}\right),
\end{align}
where we suppress the z-component which is always zero.
When this photon with wave number $k$ travels under the axion background from $t$ till $t+T$, the evolved polarization components can be calculated as
\begin{align}
&\frac{e^{-ikT}}{2}\left[
e^{-i\int^{t+T}_t \delta \omega {\rm d}t}
\left(\begin{array}{c}
1 \\
i \\
\end{array}\right)
+
e^{+i\int^{t+T}_t \delta \omega {\rm d}t}
\left(\begin{array}{c}
1 \\
-i \notag\\
\end{array}\right)\right]
\\
&=e^{-ikT}
\left(\begin{array}{c}
\cos (\int^{t+T}_t \delta \omega\, {\rm d}t  ) \\
\sin (\int^{t+T}_t \delta \omega\, {\rm d}t ) \\
\end{array}\right).
\end{align}
%
The rotation angle of the linear polarization plane is given by
\begin{align}
\theta(t,T)=\int^{t+T}_t \delta \omega(t)\, {\rm d}t
=-\frac{g_{a\gamma}}{2}\left[\phi(t+T)-\phi(t)\right],
\label{time integral}
\end{align}
where the spatial argument of the ADM field is omitted while $\phi$ should be evaluated at the both ends of the light path,
$\phi(t)=\phi(t,\bm{x}(t))$ and $\phi(t+T)=\phi(t+T,\bm{x}(t+T))$.

It is interesting to compare this polarization rotation due to the ADM with Faraday rotation. Faraday rotation refers to an astrophysical phenomenon that the linear polarization plane of a photon propagating in a magnetized plasma rotates. Its rotation angle is written as~\cite{Rybicki79}
\begin{equation}
\theta_{\rm Faraday}=\frac{2\pi e^3}{m_e^2 k^2} \int{\rm d}\bm{x} \cdot \bm{B}(x)\, n_e(x),
\label{theta Faraday}
\end{equation}
where $m_e$ is the electron mass, $n_e$ is the plasma number density,  $\bm B$ is the magnetic field sustained by the plasma,
and $\int{\rm d}\bm{x}$ denotes the line integral along the light path. If the magnetic field could be described by a potential  $\bm{B}=\bm{\nabla}\Phi$
and $n_e$ was homogeneous,
the Faraday rotation would be determined only by the boundary condition like Eq.~\eqref{time integral}. However, since $\bm{B}=\bm{\nabla}\times \bm{A}$
and $n_e$ is inhomogeneous in reality, the evaluation of $\theta_{\rm Faraday}$ depends on the light path.

The present background ADM field is well approximated by
\begin{equation}
\phi(t,\bm x) = \phi_0\cos(m t + \delta(\bm x))\,, 
\label{eq: axion}
\end{equation}
with the constant amplitude $\phi_0$, the axion mass $m$ and a phase factor $\delta(\bm x)$ whose spatial dependence will be discussed soon.
In the case of homogeneous $\delta$, we can evaluate the rotation angle as
\begin{equation}
\theta(t,T)\approx 2\times10^{-2} \sin\Xi\,\sin(mt+\Xi+\delta)\,
g_{12}\, m_{22}^{-1}\,,
\label{theta}
\end{equation}
where $g_{12}\equiv g_{a\gamma}/(10^{-12}{\rm GeV}^{-1}),\ m_{22}\equiv m/(10^{-22}{\rm eV})$ and we used the dark matter density around us, $\rho_{\rm DM} = m^2 \phi_0^2/2 \approx 0.3~\text{GeV}/\text{cm}^{3}$~\cite{Read:2014qva}.
$\Xi\equiv mT/2\approx 750(c\, T/100{\rm pc})m_{22}$ is typically much larger than unity
and hence the factor $\sin \Xi$ is expected to be $\mathcal{O}(1)$.

If the spatial dependence of the phase $\delta(\bm x)$ in Eq.~\eqref{eq: axion} is negligible as we assumed in Eq.~\eqref{theta}, the axion background coherently oscillates everywhere.
However, on larger scale than its de Broglie wavelength 
\begin{equation}
\lambda= \frac{2\pi \hbar}{m v}\approx 400 {\rm pc}\  m_{22}^{-1}v_3^{-1}\,,
\end{equation}
with $v_3\equiv v/(10^{-3})$~\cite{Lentz:2017aay}, the axion field is not expect to show a coherent oscillation. 
Even in that case, since Eq.~\eqref{time integral} still holds, only the phases at the light source $\delta(\bm{x}_{\rm src})$ and the observer $\delta(\bm{x}_{\rm obs})$ are relevant. One can show that the sine functions in Eq.~\eqref{theta} are replaced by $\sin[\Xi+(\delta_{\rm src}-\delta_{\rm obs})/2]\sin[mt+\Xi+(\delta_{\rm src}+\delta_{\rm obs})/2]$ in that case. 
Thus the evaluation in Eq.~\eqref{theta} basically remain unchanged.

Replacing the first sine function by its standard deviation for a random argument $\sin(\Xi)\simeq 1/\sqrt{2}$ because of $\Xi \gg 1$, we obtain
\begin{align}
\theta(t,T)\simeq 
&\,1.4\times 10^{-2} \sin (mt+{\rm const.}) \, g_{12}m_{22}^{-1}\,.
\label{predicted theta}
\end{align}
%
It should be stressed that since the rotation angle oscillates with the same frequency as the ADM field $\phi(t)$, one can in principle measure the axion mass from the photon polarization.
This oscillatory behavior of the signal is also advantageous
to distinguish it from other various effects.

\section{III. Polarimetry of PPDs}

\begin{figure}
\begin{center}
\includegraphics[height=6.0cm]{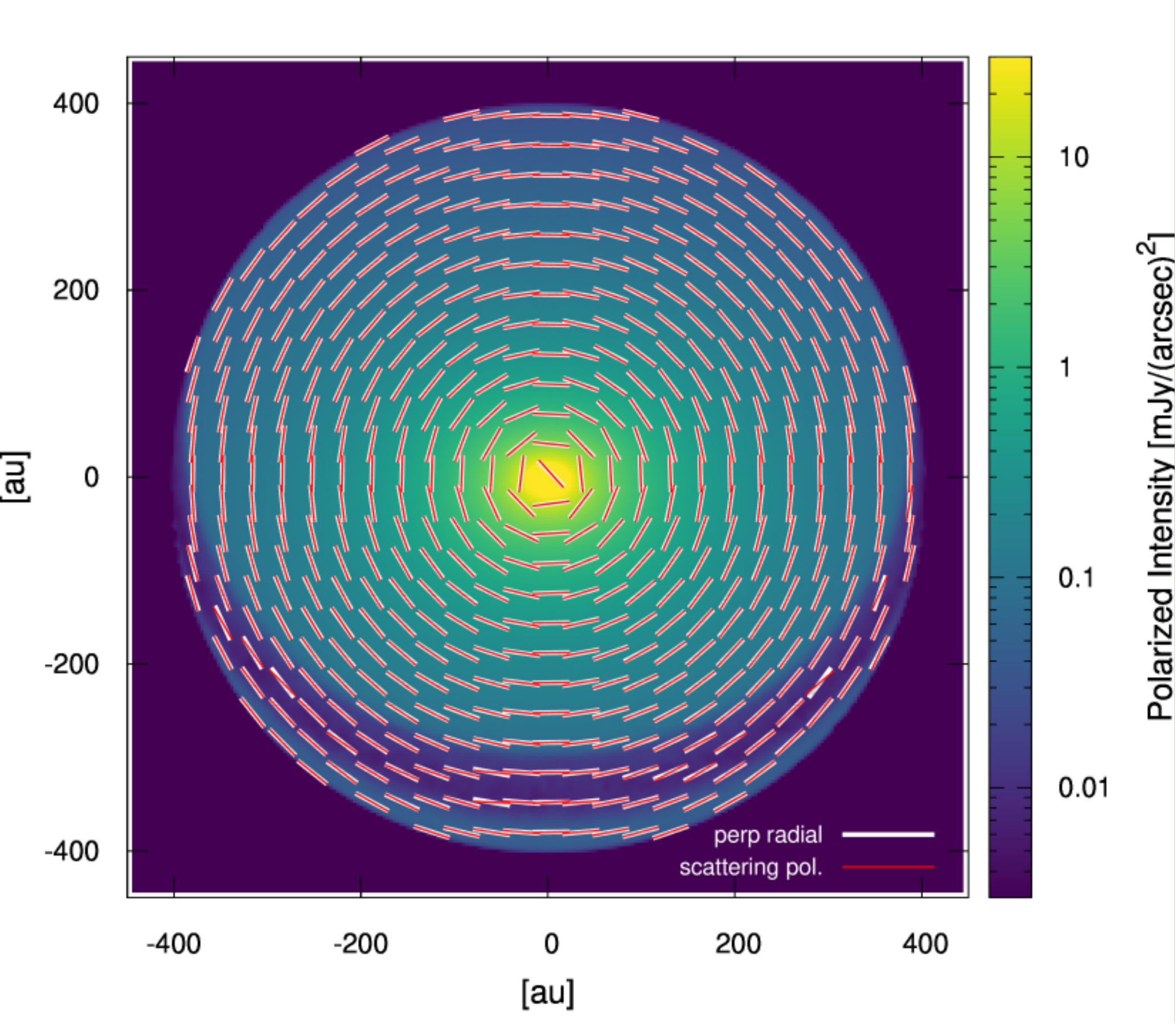}
\includegraphics[height=6.0cm]{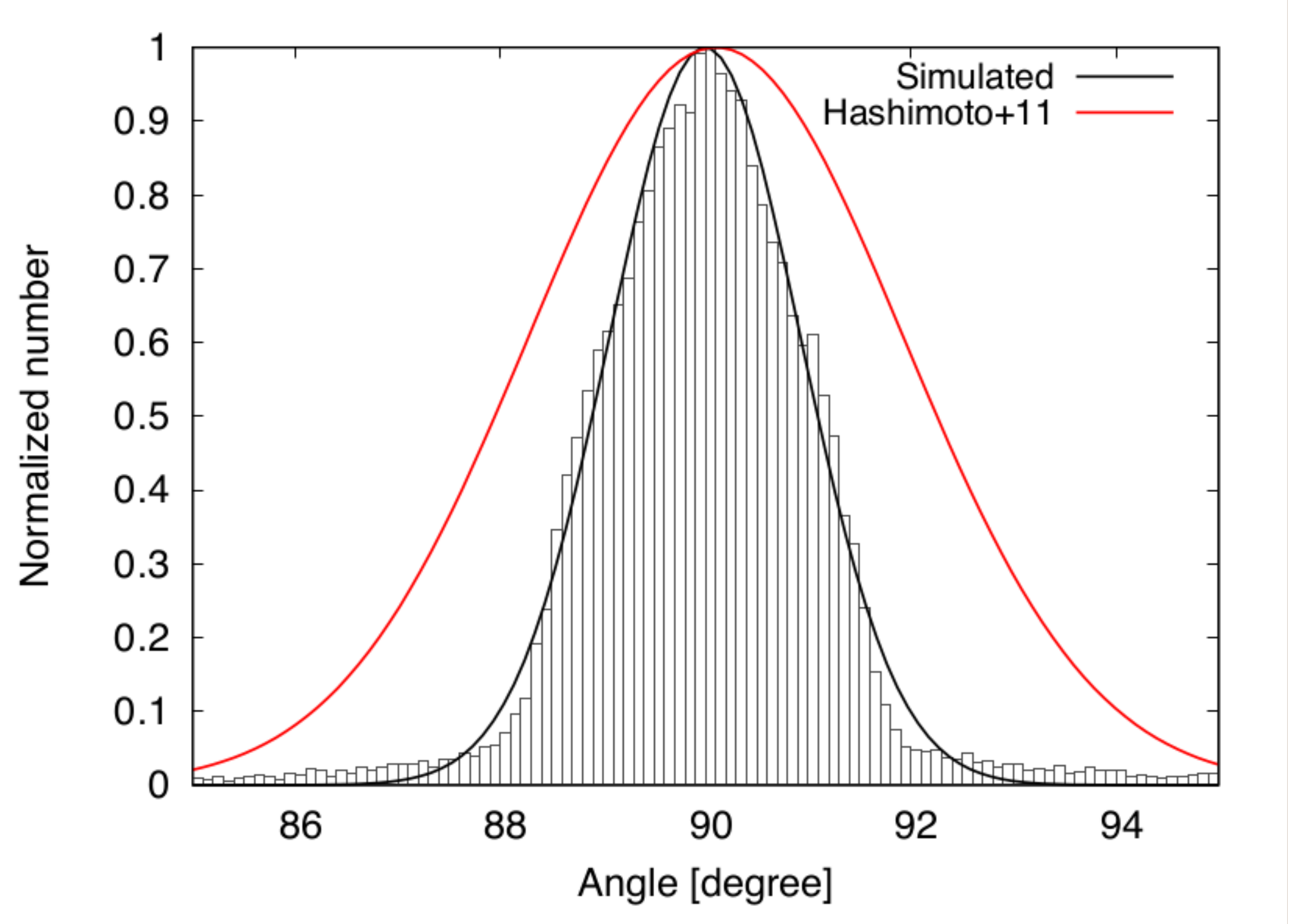}
\caption{(Top) Simulated polarization vectors of AB Aur at $\lambda=1.6\ \mu$m. Red and white bars represent the direction of scattering polarization and perpendicular to the radial vectors, respectively. (Bottom) Histogram of polarization angles with respect to radial vectors. Black and red lines are the Gaussian fits of the histograms of simulated and observed values from \citet{Hashimoto:2011xe}, respectively.}
\label{fig:ppdpol}
\end{center}
\end{figure}

Faraday rotation is well known and firmly observed for photons propagating in a magnetized plasma. 
The photon linear polarization plane can rotate even in vacuum if $CPT$ invariance is broken either spontaneously or explicitly \cite{myers03,toma12}. In these cases the rotation angle depends on the photon energy (e.g. $\theta_{\rm Faraday}\propto k^{-2}$ in Eq.~\eqref{theta Faraday}), so that one can detect this effect without knowing the intrinsic linear polarization angle at the source by comparing the angles measured at different frequencies. In contrast, $\theta(t,T)$ due to the ADM effect (Eq.~\ref{predicted theta}) does not depend on the photon energy. To detect or constrain the ADM effect, we require the intrinsic linear polarization angle at the source.

The best source for this purpose as far as we are aware is PPD.
From optical to near-infrared wavelengths, (sub)micron-sized dust particles at the disk surface scatter the central star's light. Since the disk scattered light dominates disk surface brightness in these wavelengths, polarimetric observations can provide polarization angles of scattered light \cite[e.g.][]{Hashimoto:2011xe}. Since scattered light is polarized perpendicular to the scattering plane, polarization angles become perpendicular to radial vectors from the star in observed images \cite{Murakawa2010}. For example, for a face-on disk, polarization vectors are azimuthal direction. Even if the disk is inclined toward the observer or has some physical structures such as dust density fluctuations, disk regions directly illuminated by the star will produce polarization perpendicular to the radial vectors.

For an illustration of the linear polarization pattern of PPDs, we perform a 3D Monte Carlo radiative transfer simulation. We set physical parameters that correspond to a famous PPD, AB Aur, for our simulation (e.g. the inclination angle $\sim 27$ degree),\footnote{We follow a disk and dust model of AB Aur adopted in Li et al. \cite{Li16}; however, we set maximum dust grain radius as 0.65 $\mu$m in order to reproduce observed polarization fraction at $\lambda=2\ \mu$m \cite{Perrin09}. Our 3D Monte Carlo radiative transfer simulation is performed by using a publicly available code RADMC-3D \cite{Dullemond12}. Number of photon packet used in a scattering Monte Carlo run is $10^9$.} and Figure \ref{fig:ppdpol} shows the result of model image and the histogram of polarization angles with respect to radial vectors from the central star on the sky.
We perform Gaussian fit to the histogram, and then obtain the peak value and FWHM are $89^\circ.99\pm 0^\circ.01$ and $2^\circ.19\pm0^\circ.020$, respectively.
This clearly illustrates that the intrinsic linear polarization angles of PPDs do not systematically shift from $90^\circ$.
Therefore, systematic deviation from $90^\circ$ would imply the polarization rotation effect of ADM. We should note that deviation from azimuthal polarization pattern is expected for a disk with large inclination angle due to multiple scattering \citep{Canovas15, Pohl17} or exotic grain properties \citep{Kirchschlager14}. Thus, disks with almost pole-on view, like AB Aur disk, are in favor of probing ADM.

In order to detect polarization from disks, high-spatial resolution observation is necessary because low-resolution observation depolarizes the signal. PPDs have typically $\approx 10^2$ au radii. The diffraction limit of a 8 meter telescope at $\lambda=1.6\ \mu$m is about 0.05 arcsec, which can spatially resolve 5 au structure of PPDs at distance of 100 pc. Therefore, current telescopes have enough power to spatially resolve PPDs, although atmospheric fluctuation may make angular resolution worse. In addition, disk radii are much smaller than the de Broglie wavelength; therefore, scattered light is expected to arise in the same coherent axion field.

At near-infrared wavelengths, a number of polarimetric imaging observations of PPDs have been performed. These observations have demonstrated that observed polarimetric pattern is consistent with disk scattered light interpretation \cite[e.g.][]{Hashimoto:2011xe, Momo15}. Among these, an observation of the object, AB Aur, performed by \citet{Hashimoto:2011xe} presents a useful data to assess the birefringence.
They showed that a histogram of polarization angles with respect to radial vectors can be fitted by a Gaussian function with peak value of $90^\circ.1\pm0^\circ.2$ and FWHM of $4^\circ.3+0^\circ.4$ (see the red line of the bottom panel of Figure \ref{fig:ppdpol}). Within the 1$\sigma$ accuracy, the peak value of polarization angles is consistent with that of scattered light. In other words, the systematic shift of polarization angles due to the axion field should be less than $0^\circ.3$ for 1$\sigma$, which corresponds to $5\times10^{-3}$ radian. Therefore, the current upper bound on the rotation angle $\theta$ caused by the ADM is
\begin{equation}
|\theta| <5\times 10^{-3},
\label{observed theta}
\end{equation}
where the distance to the source is 162.9pc~\cite{gaia16,gaia18}.

One caveat of the above discussion is that \citet{Hashimoto:2011xe} might overlook some systematic errors. It is possible to obtain larger errors if one analyzes the observational data by taking account of a full error budget, and hence, the above estimate might be optimistic. 
Instrumental errors on polarization angles may be different for pixels one by one in the image and can affect both the peak value and FWHM of the polarization angle distribution.
In such a case, the FWHM of the Gaussian may reflect the systematics.
If we assume that the systematics is similar to the FWHM value, $4^\circ.3$, the upper bound on the rotation angle becomes $|\theta| <7\times 10^{-2}$. We adopt this value as a conservative upper bound.

\section{IV. Constraint on coupling constant}

Comparing Eqs.~\eqref{predicted theta} and \eqref{observed theta},
we can place  upper limit on $g_{a\gamma}$. 
\begin{equation}
g_{a\gamma}<5\times10^{-13} m_{22}{\rm GeV}^{-1}\,,
\label{const on gag}
\end{equation}
where $c T=162.9$pc is substituted and $\sin (mt+{\rm const.})$ is replaced by $1/\sqrt{2}$ again. It is worth noting that total integration time of the polarized intensity image in \citet{Hashimoto:2011xe}  is 189.6 s, and hence, the oscillation of the axion field can be negligible. 

\begin{figure}
\begin{center}
\includegraphics[height=6.0cm,keepaspectratio]{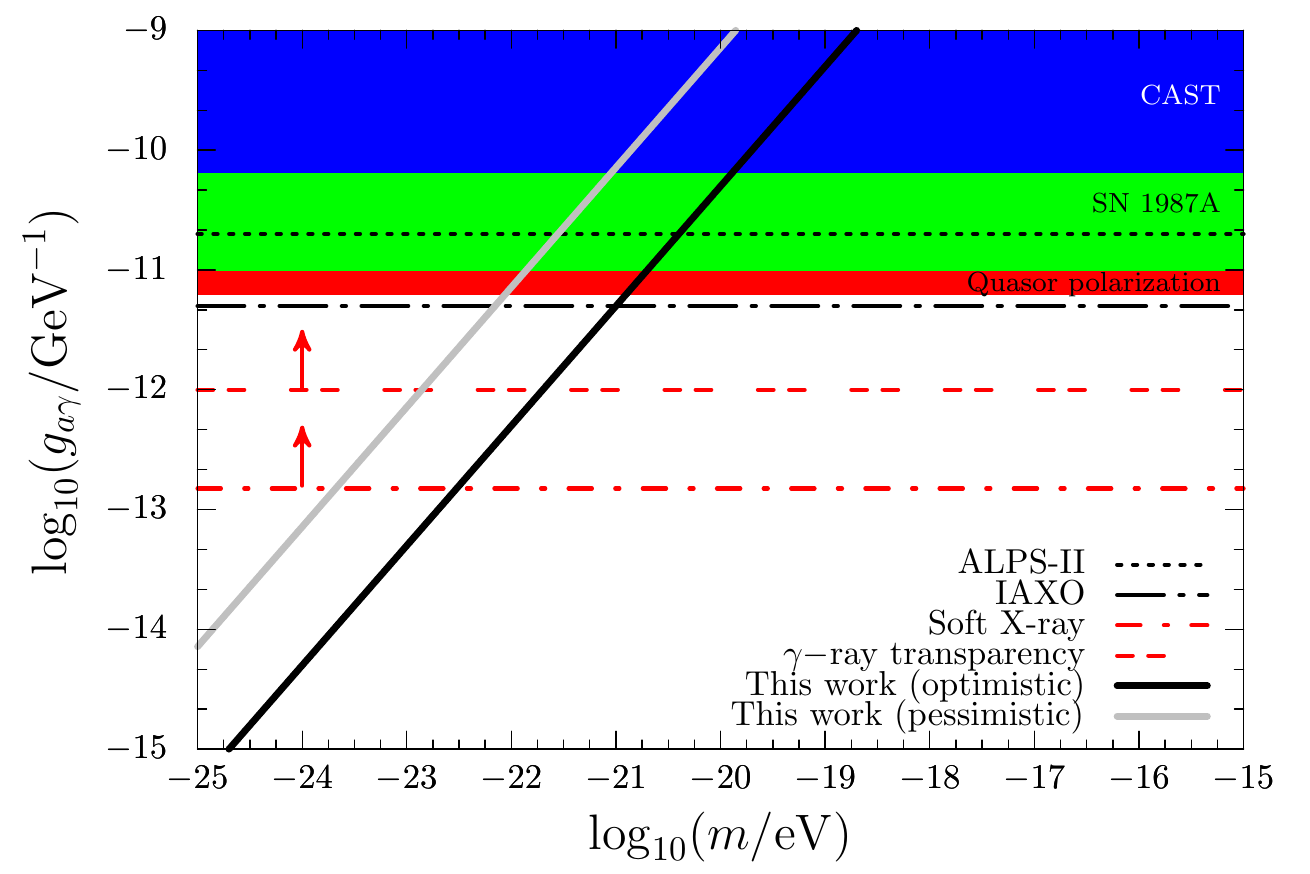}
\caption{Constraint on the coupling constant $g_{a\gamma}$ for varying mass of the axion dark matter $m$. 
The black solid line indicates the upper bound derived in eq.~\eqref{const on gag}, whereas the gray solid line is the conservative upper bound. Blue, green, and red regions are rejected by the experiment (CAST~\cite{Anastassopoulos:2017ftl}) and astronomical observations (SN 1987A~\cite{Payez:2014xsa} and Quasor polarization~\cite{Payez:2012vf}). Black broken lines are the expected sensitivities of future projects (ALPS-II~\cite{Ehret:2010mh} and IAXO~\cite{Armengaud:2014gea}) and red broken lines are presumptive lower limits from observations (Soft X-ray~\cite{Angus:2013sua} and $\gamma$-ray transparency~\cite{Meyer:2013pny}).}
\label{fig:constraint}
\end{center}
\end{figure}

In fig.~\ref{fig:constraint}, we show our new constraint with the previous bounds as well as the sensitivity curves of the future experiments
which are taken from a review paper~\cite{Dias:2014osa}.
One can see that we improve the constraint on $g_{a\gamma}$ for $m\lesssim 10^{-21}$eV.
Note that the red broken lines in fig.~\ref{fig:constraint} are the  lower bound on $g_{a\gamma}$ obtained by assuming the soft X-ray excess in the Coma cluster and the transparency of very high-energy $\gamma$-ray are caused by the axion-photon conversion~\cite{Angus:2013sua, Meyer:2013pny}.
Although these arguments are not necessarily firm against astrophysical uncertainties (e.g. intrinsic photon spectra at the sources in TeV energy range \cite{biteau15}), it is remarkable that our bound reach them for the first time in this ultra-light mass region.

\section{V. Discussion}

In this Letter, we considered the rotation of the linear polarization plane of propagating photon due to the ADM and its observation with PPDs.
Our new approach derived a stronger constraint on $g_{a\gamma}$
than the previous works for $m\lesssim 10^{-21}$eV. Although the photon birefringence caused by the ADM was also studied with the CMB polarizations, the reported bound on $g_{a\gamma}$ with a similar axion mass dependence was 2 orders magnitude weaker than ours~\cite{Liu:2016dcg}.
It implies that PPDs are suitable observation targets to search for
the ultra-light ADM, while the connection between axion and PPDs 
have never been considered to the best of our knowledge.

Furthermore, as we saw in Eq.~\eqref{predicted theta}, the rotation angle $\theta$ is predicted to oscillate with period $\sim 1.3\; m_{22}^{-1}\;$yr.
If one continuously observes the polarized light from a PPD for a long time $t_{\rm obs}\gtrsim 1.3\; m_{22}^{-1}\;$yr, the oscillation of the angle $\theta$
may be seen and it can be a smoking-gun evidence for the ADM.
This property should be useful to distinguish the ADM signature from 
the other potential effects which also modify the polarization pattern of PPDs.
It should be noted that the distance between source and earth $L=c\,T$ varies in time due to the relative motion and it might distort the predicted oscillatory behavior of $\theta$. However, we can measure the relative motion through the Doppler effect and correct its influence.

While the rotation angle of the photon linear polarization plane highly depends on the photon energy and the propagation distance for the Faraday effect under magnetized plasmas and for the $CPT$-invariance violation effect \cite{myers03,toma12}, those dependences are quite weak for the ADM effect as we showed above. Furthermore, the polarization of purely scattered radiation in PPDs can be observed in only optical and near-infrared wavebands, and most of well-observed PPDs are clustered in several star-forming regions in our Galaxy so that they are at similar distances. Then in order to put more stringent constraint on (or detect) the ADM effect with our method, one should keep increasing the sensitivity of polarimetic measurements of PPDs. In particular, detailed analysis of polarization angles considering full error budget is critical, which is becoming available by the state-of-the-art observations and data reduction techniques. Even re-analyses of the current polarization angle data for well observed PPDs may also be useful.

\section{Acknowledgement}
\label{Acknowledgement}

We are grateful to Atsushi Naruko for organizing a workshop `Essential next steps for gravity and cosmology' held on June 18-20, 2018 at FRIS, Tohoku University, where we came up with the idea for this Letter.
We would like to thank Pierre Fleury, Hajime Fukushima, Mikhail M. Ivanov, Dmitry Levkov, Takahiro Tanaka, Jun Hashimoto, Daisuke Toyouchi, Sergey Troitsky and Ken'ichi Saikawa for useful discussions and comments.
This work is partially supported by the Grants-in-Aid for JSPS Research Fellow No.~17J09103 (T.F.) and No.~17J02411 (R.T.), and by JSPS Grants-in-Aid for Scientific Research 18H01245 (K.T.).

\end{document}